\documentclass{pasj00}

\begin{document}
\SetRunningHead{Deguchi et al.}{Maser Emission in the Infrared Dark Cloud against the Galactic Center}
\Received{2011/04/26}
\Accepted{2011/10/11}

\title{Maser Emission toward the Infrared Dark Cloud G359.94+0.17 Seen in Silhouette against the Galactic Center}

\author{Shuji \textsc{Deguchi}, }
\affil{Nobeyama Radio Observatory, National Astronomical Observatory, \\
and Department of Astronomical Science, The Graduate University for Advanced Studies, \\
Minamimaki, Minamisaku, Nagano 384-1305}
\author{Daniel \textsc{Tafoya},} 
\affil{Department of Physics and Astronomy, Graduate School of Science and Engineering, \\ 
Kagoshima University, 1-21-35 Korimoto, Kagoshima 890-0065}
 \and
 \author{Nagisa \textsc{Shino}}
\affil{Graduate School of Science and Engineering, Yamaguchi University, \\
1677-1 Yoshida, Yamaguchi, Yamaguchi 753-8512} 
\KeyWords{masers ---  ISM: clouds ---
ISM: dust, extinction  -- 	ISM: molecules}  

\maketitle

\begin{abstract}
The infrared dark  cloud G359.94+0.17 is a conspicuous, opaque cloud, 
which is seen in silhouette against the Galactic center.  
We found   unexpectedly strong  ($\sim 50$ Jy) maser emission of CH$_3$OH at 44 GHz with additional
weak 22 GHz H$_2$O maser and 43 GHz SiO thermal emissions  toward this cloud.
Detections of these molecular  lines indicate that strong  star forming activities are proceeding  in this cloud,
 which were not reported previously despite of numerous works toward the Galactic center.
The line profiles of the NH$_3$ inversion lines at 23 GHz  indicate that G359.94+0.17
is composed of mainly two clouds with $V_{\rm lsr}= 0,$ and 15  km s$^{-1}$ overlapped  
on the line of sight. The maser emission is associated with the 15 km  s$^{-1}$ cloud, 
suggesting that it is located at the Norma spiral arm. 
\end{abstract}

\section{Introduction\label{sec:Intro}}
One of the infrared dark clouds seen in projection to the Galactic center, G359.94+0.17 \citep{dut02}, is
the most distinctive dark cloud seen within $0.3^{\circ}$ of the Galactic center.
(In Figure 1,  we show a part of  the Spitzer/GLIMPSE false-color image\footnote{available at 
http://apod.nasa.gov/apod/fap/image/0906\\ /mwcenter\_spitzer\_big.jpg ;
The thick dark patch is also clearly seen in  the 2MASS  and MSX false-color images
as well as on WISE (NASA's Wide-field Infrared Survey Explorer mission) images
at various infrared wavelengths.}
in this direction for an illustration purpose; \cite{sto06}.)
In the area of a size of a few arc minutes toward this cloud, almost no star in the Galactic bulge is seen.
This cloud causes substantial undersampling of AGB stars in the Galactic bulge in this direction 
(for example, see figure 1 of \cite{deg04}). 

Though numerous studies of molecular clouds toward the Galactic center have been made in the past
(for example, see \cite{mor96,yus09}), 
a very limited amount of works were made to investigate this interesting cloud G359.94+0.17, or 
more in general, the dark clouds toward the Galactic center. This is
possibly because these clouds are supposed to be in the foreground and not directly associated with the 
Galactic center.
However, without cautious studies, it is hasty to decide the distances to these clouds.

\citet{nag09} carefully studied the  extinction due to dust grains  in the $K_S$ band within the $5^{\circ} \times 2^{\circ}$ area in this direction.
They found that the $V_{\rm lsr}=15$ -- 20 km s$^{-1}$ CO emission is associated with these clouds. Based on the  star counts in the $JHK$ bands,
they estimated the distances to these clouds to be 3.2 -- 4.2 kpc, and suggested that  these clouds are located in the Norma spiral arm.
The dark clouds that they studied have extinction of $A_K=0.6$--$1.0$.  Though their contour map (figure 2 of \cite{nag09})
indicated that the $V_{\rm lsr}=15$ -- 20 km s$^{-1}$ CO emission is also associated with the dark cloud G359.94+0.17,
they excluded  this cloud from their study, possibly because $K$-band extinction of this dark cloud is
too high for their star-count study. 

 In this paper, we report the 
 detection of  maser emission of  CH$_3$OH and H$_2$O
toward the dark cloud, G359.94+0.17. The presence of the maser emission toward this dark cloud 
suggests that strong star forming activities are proceeding in this infrared opaque cloud. 
We have also observed thermal emission of SiO and NH$_3$. Based on these observations,
we discuss the nature of this interesting dark cloud which is seen in projection to  the Galactic center. 

\begin{figure}
\begin{center}
\FigureFile(70mm,70mm){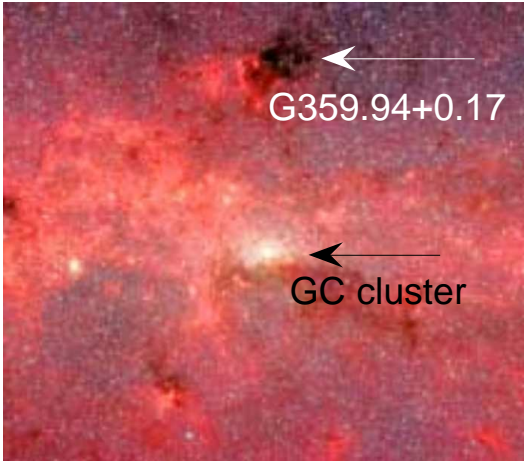}
\end{center}
\caption{A part of the Spitzer false-color image (3.6--8 $\mu$m) of the Galactic center region 
($0.55 ^{\circ}\times 0.5^{\circ}$; left is the direction of increasing the Galactic longitude).
White arrow indicates the dark cloud G359.94+0.17 and black arrow indicate the Galactic center star cluster
surrounding Sgr A* at ($l$, $b$)=(359.944$ ^{\circ}$,$-0.046 ^{\circ}$). 
\label{fig:GC2}
}
\end{figure}
\begin{figure}
\begin{center}
\FigureFile(70mm,120mm){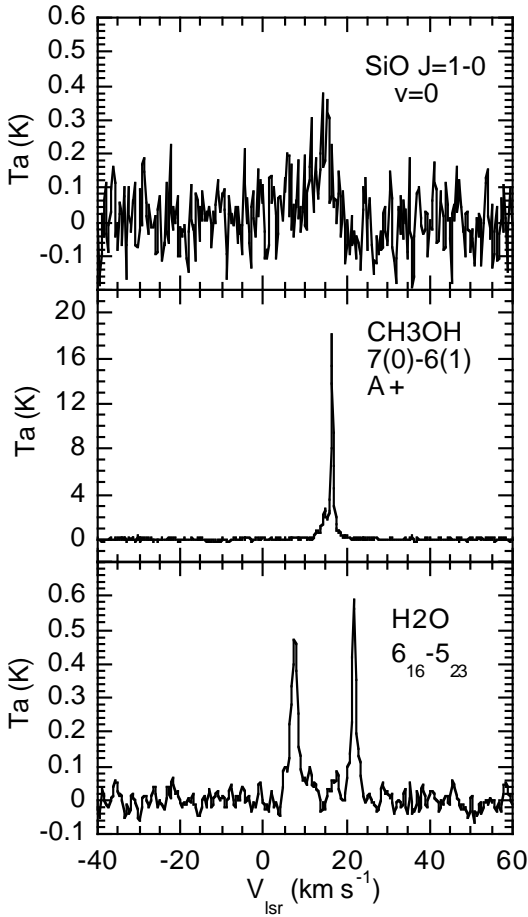}
\end{center}
\caption{Spectra of the SiO $J=1$--$0, v=0$ (top),  CH$_3$OH 7(0)--6(1) A+ (middle), and
H$_2$O $6_{16}$--$5_{23}$  (bottom) transitions toward the dark cloud G359.94+0.17 (H$_2$O; 
17h44m53.20s,  $-28^{\circ}$54$'$04.8$''$).
\label{fig:2}
}
\end{figure}

\section{Observations and Results\label{sec:Obs}}
Observations of the dark cloud G359.94+0.17   were made on 2011 February 26  with the Nobeyama 45-m telescope
in the H$_2$O and NH$_3$ lines between 22 and 24 GHz, and on March 10 in the SiO and CH$_3$OH lines 
between 42 and 44  GHz.  
The half-power full beam width (HPFBW) was about  70$''$ at 22 GHz and 40$''$ at 43 GHz.
We used cooled HEMT receivers, H22 and H40 ($T_{\rm sys} \sim$ 100--250 K)
and acousto-optical spectrometer arrays with high (40 kHz; AOS-H) and
low (250 kHz; AOS-W) resolutions having 2048 channels each.
The conversion factor of the antenna temperature ($\equiv T_a^*$) to the flux density was
$\sim 2.9$ Jy K$^{-1}$  at 43 GHz  and it is 2.8 Jy K$^{-1}$  at 22 GHz.  The accuracy of pointing of the telescope was
about 5$''$.
The observational results for  the CH$_3$OH, SiO, and H$_2$O lines are summarized in table 1 and their spectra are shown in Figure 2. 

\begin{table*}
  \caption{Observed line intensities of the CH$_3$OH,  SiO,  and H$_2$O lines.}\label{tab:1}
 \begin{center}
  \begin{tabular}{lccccccr}
  \hline\hline 
 Frequency & Molecule   & Transition    &  $T_a$(peak) & $V_{\rm lsr}$(peak)  &  Integ. Ints. &  r.m.s \\
    (GHz)      &                   &                      & (K) & (km~s$^{-1}$) & (K~km~s$^{-1}$) & (K)\\
\hline
 44.06949   &  CH$_3$OH  &  7(0)--6(1) A+   &  18.029 & 16.5 & 19.245 & 0.087 \\
 43.423858 & SiO                &   1--0  $v=0$      &   0.409  &  15.3 &  1.648 & 0.087 \\
 22.235077 & H$_2$O       &   6(1,6)--5(2,3) & 0.485 &   7.6  &  1.459 & 0.024 \\
 22.235077 & H$_2$O       &   6(1,6)--5(2,3) & 0.606 & 21.6  & 1.295 & 0.024 \\
   \hline
  \end{tabular}
  \end{center}
\end{table*} 

The H$_2$O maser emission consists of two components at $V_{\rm lsr}=7.6$ and 21.6 km s$^{-1}$.
To check the positions of  H$_2$O maser emission,  we have made  the  5-point 40$''$-grid cross mapping observation as follows.
From the 2MASS $K$-band  image, we found with eye inspection that the position of the darkest part of 
the cloud is (R.A., Decl.)=(17h44m54.0s,  $-28^{\circ}$54$'$00$''$) in epoch of $J2000$.
After detecting H$_2$O masers at this position, we have performed the 5-point 40$''$-grid cross mapping of 
H$_2$O maser emission. 
From the integrated intensities of  the H$_2$O maser components at each position of the grid, we computed
the center positions of each water maser component, assuming that they are point sources.
The resulted positions are  (R.A., Decl.)=(17h44m53.65s,  $-28^{\circ}$54$'$04.1$''$) for the $V_{\rm lsr}=7.6$ km s$^{-1}$ component,
and  (R.A., Decl.)=(17h44m53.20s,  $-28^{\circ}$54$'$04.8$''$) for the $V_{\rm lsr}=21.6$ km s$^{-1}$ component in epoch of $J2000$,
which corresponds to ($l$, $b$)=(359.947$^{\circ}$, +0.154$^{\circ}$), and ($l$, $b$)=(359.946$^{\circ}$, +0.155$^{\circ}$), respectively.
It is known that the accuracy of pointing  for the 45m telescope is about 5$''$ in the absolute position.  Therefore, we assume that
the water maser position given above is accurate at this level.
These positions are approximately $1'$ east of the position of the dark cloud G359.94+0.17 given by \citet{dut02}.

The mapping observations were also made for the NH$_3$ inversion lines simultaneously with the mapping for the H$_2$O maser line.
The variations of the NH$_3$ 1(1) and 3(3) line profiles at $40''$ separated positions are shown
in Figure 3. Due to the hyperfine splitting, 
the  NH$_3$  1(1) line  is separated into 5 groups with velocity separations of approximately  $-19$, $-8$,  8, 19 km s$^{-1}$ 
relative to the strongest central components \citep{ryd77}.  
Left panel of Figure 3 shows that  the NH$_3$  1(1) line profile is composed of 7 peaks, which suggests that the cloud has several different velocity components.
The strongest peak of the NH$_3$ lines falls at $V_{\rm lsr}\sim 15.2$ km s$^{-1}$
at position (00$''$, 00$''$)  [= (17h44m54.0s,  $-28^{\circ}$54$'$00$''$)] and the second strongest peak
at $V_{\rm lsr}\sim 0.3$ km s$^{-1}$ which exhibits stronger emission at position ($-40''$, 0$''$).  
In addition we found  a weak enhancement of emission at $V_{\rm lsr}\sim 80$ km s$^{-1}$  at position (+40$''$, 00$''$),
which is indicated by arrows in figure 3. 
This emission exhibits  a broad line width of more that 30 km s$^{-1}$. 

\section{Discussions\label{sec:disc}}

A  large number of infrared dark clouds have been cataloged in the first and forth quadrants of
the Galactic plane using the 8.3 $\mu$m MSX images  \citep{sim06a} 
and the Spitzer GLIMPSE and MIPS data \citep{per09}. 
Their nature has  been well studied through molecular lines \citep{sim06b}.  
The IR dark clouds contain compact cores with a mass range 10 -- 2100 $M_{\odot}$ \citep{rat06},
 and smaller flagments with masses down to $\sim 0.1$ $M_{\odot}$ \citep{per10}.
 Approximately one third of these cores show evidence of active star formation \citep{jac08b}.
About 12\% of these infrared dark clouds exhibit H$_2$O maser emission \citep{wan06}.
 It seems that  the dark cloud G359.94+0.17 shares similar properties  in several respects 
 as the known IR dark clouds except for its massive feature. 
It is listed  as the dark cloud core G359.91+00.17a  [ ($l$, $b$)=(359.938, +0.174)]
in the catalog of \citet{sim06a}, where "a" indicates the multiplicity index code 
in the dark cloud complex G359.91+00.17.

\begin{figure*}
\begin{center}
\FigureFile(160mm,130mm){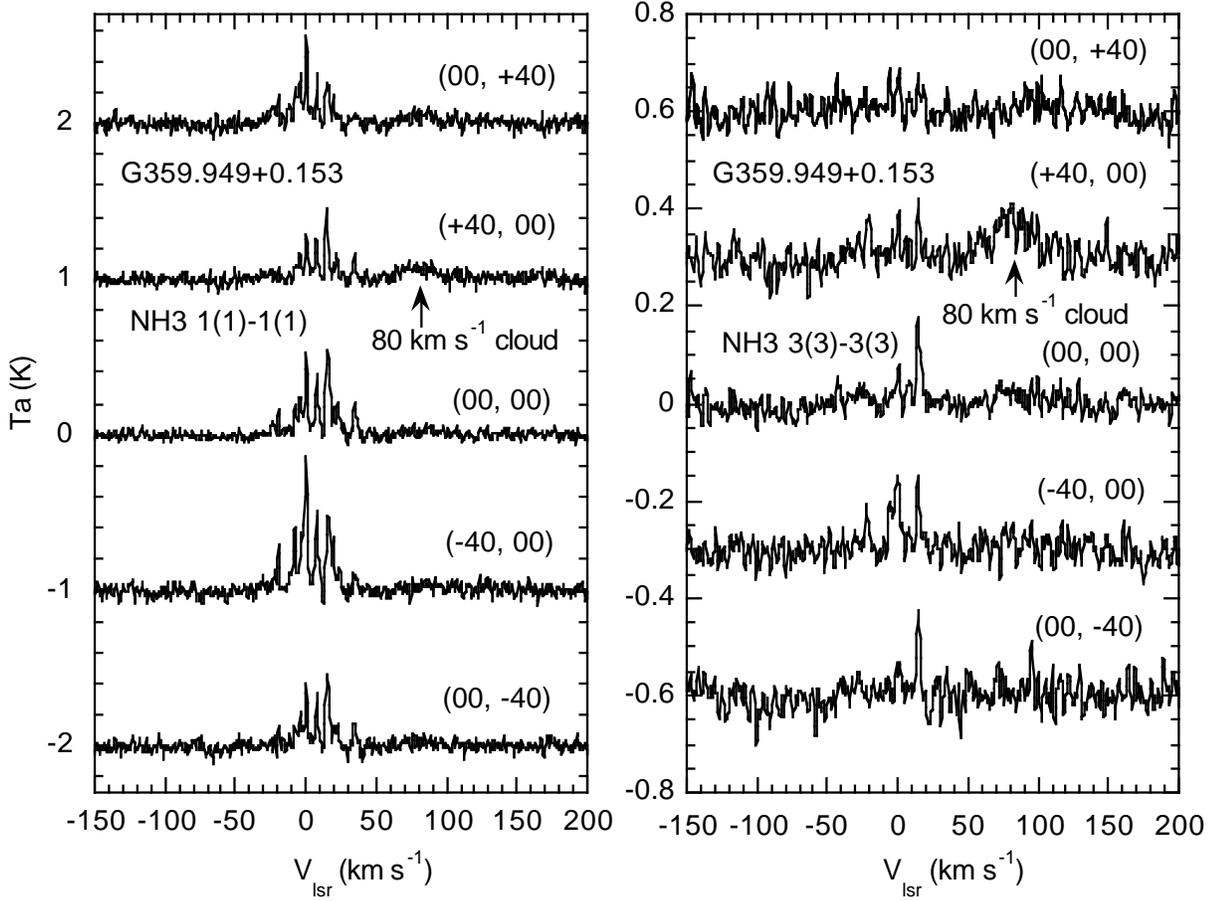}
\end{center}
\caption{5-point 40$''$-grid cross mapping spectra of the NH$_3$ 1(1) (left) and 3(3)  (right) lines.  
The center velocity of the 1(1) line refers to 23.69451 GHz.  
The mapping center (0,0) is  the position of (R.A.=17h44m54.0s, Decl.= $-28^{\circ}54' 00''$; J2000),
and the position shifts are made  in the R.A. and Decl. directions.  In order to reduce noises, three-channel running mean
is applied for the NH$_3$ 3(3) spectra. The emission around $V_{\rm lsr}=80$ km s$^{-1}$,  which is indicated by the arrow,  
is detected at $+40''$ east more clearly in the 3(3) line. The influence of the hypersplittings is marginal 
in the 3(3) spectra at this noise level.
\label{fig:3}
}
\end{figure*}


\subsection{Thermal emission of NH$_3$ and SiO \label{sec:velocity}}
 The intensity ratio between the NH$_3$ inversion lines can be used as a thermometer of the cloud
\citep{dan88}. 
 We calculated the rotational temperature of the observed clouds from the 1(1) and 2(2) line intensity ratio
 and summarized the result in table 2.  
We derived the rotational temperature from the 1(1) versus 2(2) line intensity ratio at each position and gave the result in table 2. 
They are in a range between 19 and 25 K.
The intensity of the NH$_3$ sub-component at $V_{\rm lsr}\sim 80$ km s$^{-1}$ is stronger at 40$''$ east of the mapping center.
The absence of the hyperfine feature of this component in the 1(1) profile indicates that this cloud has large internal motion, 
which flattens out the hyperfine feature.
In contrast, the hyperfine feature of the  components appearing 
at $V_{\rm lsr}\sim -10$ -- +30 km s$^{-1}$  is clearly seen in right panel of Figure 3, 
which indicates that internal motions of these clouds are less than a few km s$^{-1}$.
It exhibits  a complex variation in the hyperfine feature at different positions. 
We identified that the major components are two clouds with $V_{\rm lsr}=-0.3$ and 15.3 km s$^{-1}$,
which are easily seen in the 3(3) spectra, 
because the influence of the hyperfine splittings is marginal 
in the 3(3) spectra at this noise level.\footnote{
Because the hyperfine intensities of the satellite components  in the 3(3) line
are negligibly small  ($\sim 10$ \% in total ; \cite{cou06}), 
the line can be regarded as a single component at the noise level in the present paper.
} %
We named them as primary and  secondary components. In addition, we call the 80 km s$^{-1}$ component as  tertiary. 
The intensity of these three components varies depending on the positions. 
Because the grid separation 40$''$ is just slightly larger than 
the half-power half-beam width (HPHBW $\sim 35''$) of the telescope, and because the hyperfine lines
from the different components of the cloud are blended in the line profile, 
it is cumbersome to clearly  separate all the velocity components from the observed profile in this observation.   
We computed a contribution of each component by successive decomposition of the original line profile into  the standard NH$_3$ profiles with hyperfine splitting,
and gave these numbers in the foot of table 2. The profiles shown in Figure 3 are well fitted by introducing
two major components with $V_{\rm lsr}=0.3$ and 15.2 km s$^{-1}$ (with  additional very minor contributions ($<10$\%) from 
the $-3.5$ and $-24.5$  km s$^{-1}$ components.)

We also computed the column density of the NH$_3$ at the 1(1)  levels 
[see details in \citet{lis06}] and  listed it in the last column  in table 2. 
The column density gives the optical extinction 
$A_v=35$ and 28 at the position ($-40''$, $0''$) and ($0''$, $0''$), respectively, and
  $A_v=13$ for tertiary component at the position  ($+40''$, $0''$). 
Here we have assumed that the NH$_3$ abundance is $2\times 10^{-9}$ per H$_2$ \citep{lis06},
and that the level populations of  ammonia are thermalized at the observed rotational temperature, and the relation between
 hydrogen column density and optical extinction as  $N_H=2.2\times 10^{21}$ cm$^{-2}\  A_v$ \citep{guv09}.
 Though it involves an uncertainty factor of 3  in the NH$_3$ abundance for these components of the cloud,
the calculated extinction $A_v\sim 30 $ ($A_K\sim 2 $) is roughly consistent with the extinction estimated in this direction
\citep{sch98}. 
The mass of the cloud can be estimated from the formula 
$M=372\  M_{\odot} \times (A_v/30)\times (D/4\ {\rm kpc})^2 \times (\theta /1')^2$,
 which assumes  a uniform cloud with a spherical shape ( i.e., a depth  being  the  same as a diameter of the cloud) 
using the above conversion ratio of $A_v$ to $N_H$, 
where $D$ and $\theta$ are the distance and angular diameter of the cloud  [cf., eq. (10) of \citet{mar09}]
and $1'= 1.16$ pc at 4 kpc. 
Approximately these clouds contain a mass of 100--500 $M_{\odot}$ in an area with an angular size of  $1'$.

\begin{table*}
  \caption{Observed line intensity of the NH$_3$ inversion lines and rotational temperature.}\label{tab:2}
 \begin{center}
  \begin{tabular}{lccccccc}
  \hline\hline 
  component   & position$^{\sharp}$ & I.I. 1(1)  & I.I. 2(2)  &  I.I. 3(3) &  $T_R$ ([1/2])  & $N_c$[NH$_3$(1,1)]\\ 
                         & ( $''$,  $''$) &(K~km~s$^{-1}$) &(K~km~s$^{-1}$) &(K~km~s$^{-1}$) & (K) & $10^{13}$ cm$^{-2}$  \\
\hline
primary  (+secondary)$^{\dagger}$ &  ($-40$,0)  & 9.147  & 3.429 & 1.832 & 24.4 & 1.7 \\ %
secondary (+primary)$^{\ddag}$ & (0,0) & 7.082 & 2.259 & 2.233 & 22.8 & 1.4 \\ 
tertiary  ($V_{\rm lsr}=$80 km s$^{-1}$)  & (+40,0) & 2.792 & 0.647 & 2.767 & 19.4 & 0.7 \\ 
   \hline
\multicolumn{7}{l}{$^{\sharp}$ (R.A., Decl.) position shift from (17h44m54.0s, $-28^{\circ}54'00''$) in J2000.}\\
\multicolumn{7}{l}{$^{\dagger}$ approximately 28\% of emission comes from the secondary (15.2 km s$^{-1}$) component.}\\   
\multicolumn{7}{l}{$^{\ddag}$ approximately 47\% of emission comes from  the primary ($ -3.5$ -- 0.3  km s$^{-1}$) component.}\\
  \end{tabular}
  \end{center}
\end{table*} 
We also detected the SiO $J=1$--0 $v=0$ thermal line at 43.4 GHz. SiO thermal emission is an indicator of an outflow
in a star forming region \citep{dow82}.  Thermal SiO emission ($J=1$--0 $v=0$) toward
the Galactic center  has been mapped  using the Yebes 14m telescope \citep{mar00}, but no emission was seen 
in their map at the position of the dark cloud G359.94+0.17.  

\subsection{H$_2$O and CH$_3$OH maser emission \label{sec:cma}}

 Methanol masers are well known tracers of early stage of star formation. They are often classified into two categories
 -- Class I and Class II  \citep{men91}. The former shows maser action mainly in the 36 and 44 GHz lines, and the latter in the 6.7 and 12 GHz lines.
 Though this dichotomy for methanol maser sources is widely used,  
 real objects occasionally exhibit confusing characteristics in high and low-mass star forming regions \citep{ell05,fon10}.
  
  Our detection of strong ($\sim 50$ Jy) 44 GHz CH$_3$OH emission  from the $V_{\rm lsr}=15$ km s$^{-1}$ component cloud of  G359.94+0.17
 suggests that this cloud is categorized to a Class I  source. 
  \citet{cas96} made an unbiased survey for 6.7 GHz CH$_3$OH emission over a 2-square-degree area toward the Galactic center
  and found 23 maser sites. However, no emission was found toward G359.94+0.17. 
  More recently, \citet{cha11} detected weak 6.7 GHz CH$_3$OH maser emission at $V_{\rm lsr}=-0.8$ km s$^{-1}$
  toward their "g22" cloud ($G359.939+0.170$) with the EVLA.
  The position of this  6.7 GHz CH$_3$OH maser was 1.5$'$ west of our  H$_2$O maser position 
  and the radial velocity indicates that  it is associated with the primary ($V_{\rm lsr}\sim 0.3$  km s$^{-1}$) component of the cloud.
 \citet{tay93} made an unbiased survey of the H$_2$O maser sources in the inner $4^{\circ}\times 4^{\circ}$ area of the Galactic center with the VLA.
 They detected water maser emission with $V_{\rm lsr}=-8.4$ km s$^{-1}$ at the position ($l$, $b$)=(359.977, 0.168), 
 which is 1.8$'$ away from our H$_2$O maser position. Positions of these masers are shown in Figure 4.
 These previously found maser sources are not related with our 44 GHz CH$_3$OH cloud (the secondary component). 
 
\begin{figure}
\begin{center}
\FigureFile(70mm,40mm){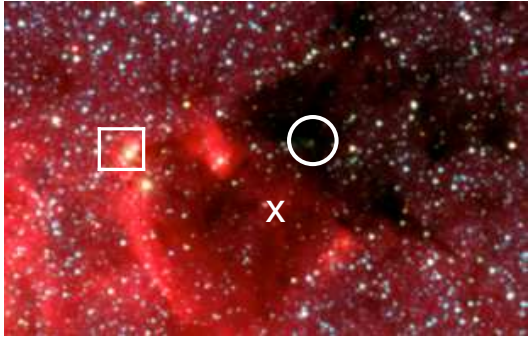}
\end{center}
\caption{Spitzer false-color image of the G359.94+0.17 region 
($7' \times 4'$; left is the direction of increasing the Galactic longitude).
Cross, circle, and square indicate the positions of the H$_2$O and CH$_3$OH masers ($V_{lsr}=7$ -- 21 km s$^{-1}$) found in this paper,  
the 6 GHz CH$_3$OH maser found by \citet{cha11} ($V_{lsr}=-0.8$ km s$^{-1}$), 
and the H$_2$O maser found by \citet{tay93}  ($V_{lsr}=-8.4$ km s$^{-1}$).
\label{fig:G359}
}
\end{figure}

Class I line surveys in the past showed that giant molecular clouds exhibits strong 44 GHz emission
with a peak intensity of typically $\sim 100$ Jy  (e.g., see \cite{has90,pra08}).  However,
the surveys toward dark clouds, specially for EGOs (Extended Green Objects: \cite{cyg09,cha11})
and for low-mass  objects \citep{bae11},  showed that  the 44 GHz  CH$_3$OH emission is weak ($\sim 1$ -- 10 Jy) in general.
Our detection of the strong (~50 Jy) 44 GHz  CH$_3$OH maser emission in the dark cloud G359.94+0.17, 
together with the detections of H$_2$O and SiO emissions, 
 suggests that  a strong outflow activity comparable with those in massive molecular clouds is proceeding in this dark cloud.

 \subsection{Other molecular lines and  distances  \label{sec:other}}
 \citet{jac08a}  
 detected CS $J=2$--1 line emission toward  ($l$, $b$)=(359.94, +0.17) (assigned as MSXDC G359.91+00.17a)
 at $V_{\rm lsr}=-0.9$ and 15.2 km s$^{-1}$ (their table 1), which agree well with the radial velocities 
 of our NH$_3$ primary and  secondary components of G359.94+0.17.  
 \citet{car98} studied the IR dark cloud G359.94+0.17 in the H$_2$CO lines at 134 -- 219 GHz, detecting the H$_2^{12}$CO and  H$_2^{13}$CO
 at  $V_{\rm lsr}=-0.2$ km s$^{-1}$. It is likely that this emission comes from the same cloud emitting the $-0.9$ km s$^{-1}$ component of CS.
  
 We found that the infrared dark cloud G359.94+0.17 is composed of three components with radial velocities,
0, 15 and 80 km s$^{-1}$.  
They may be associated with different foreground spiral arms, or with the Galactic center.
\citet{nag09} estimated the distance of the 15--20 km $^{-1}$ dark clouds toward the Galactic center as 4 kpc 
 from their star-count curve and the velocity gradient along the Galactic longitude. In addition,  6 dark clouds observed
in  6 GHz and 44 GHz CH$_3$OH masers toward the Galactic center exhibit radial velocities
in the range of 10 -- 25   km s$^{-1}$ \citep{cha11}. These facts indicate that they are in the same foreground spiral arm.
 Because the radial velocity of the secondary component of G359.94+0.17 is close to the above group of dark clouds,
 we infer that the secondary component has a similar distance as that of this group, i.e. $\sim 4$ kpc. 
 We also infer from the  large line width  ($\sim$ 30 km s$^{-1}$) and a high radial velocity 
 ($\sim 80$ km s$^{-1}$) of the tertiary component
 that the tertiary component is a cloud close to the Galactic center  (e.g., see \cite{oka10}). 
 Because  the tertiary component is seen only near the edge of the dark cloud G359.94+0.17,
and because it is located near the Galactic center,   this component is not a main source of opacity of G359.94+0.17.
This 80 km s$^{-1}$ cloud is clearly revealed in the map of the Galactic center region 
in the velocity range,  70  km s$^{-1}$ $<V_{\rm lsr}<120$ km s$^{-1}$,  in the  CS $J=1$--0 line \citep{tsu99},
 where our observed position  ($l=359.954^{\circ}$ and $b=0.144^{\circ}$) is just at the edge of this cloud
in the CS map.
 
 The distance to the primary component ($V_{\rm lsr}\sim 0$ km s$^{-1}$) is difficult to guess.  
A very  few 6 GHz CH$_3$OH masers exhibit nearly zero radial velocity toward the Galactic center:
G359.138+0.031, G0.496+0.188, and G0.836+0.184 \citep{cas10}, and  G359.199+0.041
(g31; \cite{cha11}). All of them are located at positive Galactic latitude.  
The velocity gradient of $\sim 5$ km s$^{-1}$ per degree may be derived from these sources.
The large velocity gradient \citep{sof06} indicates that they are likely associated with the Scutum-Crux arm
at a distance of $\sim 2$ kpc away from the Sun (at $l=0^{\circ}$).

\section{Conclusion}
We have detected maser emission of CH$_3$OH and H$_2$O from the infrared dark cloud
G359.94+0.17, which is seen in silhouette against the Galactic center on the near-infrared images. 
Because no star forming activity was reported  in this cloud  previously,  it was thought 
that this was a tranquil cloud at low temperature. Our observations showed that
three clouds at different distances are overlapped on the line of sight, and that this overlap of the clouds 
makes a very deep dark feature in silhouette against the Galactic center.
Detections of CH$_3$OH  and H$_2$O masers  indicate that active star formation is proceeding in one of these clouds. 
SiO thermal emission also suggests a presence of mass outflow from young stars. 
The radial velocity of this cloud ($V_{\rm lsr}\sim 15$ km s$^{-1}$) is similar 
to the velocities of a chain of dark clouds which are seen at negative latitudes in the Galactic coordinates.
It suggests that this cloud is likely located in the Norma spiral arm. It is inferred that the other component at $V_{\rm lsr}\sim 0$ km s$^{-1}$
 is slightly closer, possibly associated with the Scutum-Crux arm. 
 Though the present observation is limited only toward the direction of one distinctive dark cloud G359.47+0.17,
 it implies that diffuse clouds in the same spiral arm are widely spread in a larger ($\sim 1^{\circ}$) region toward the Galactic center.  
The foreground clouds must seriously  contaminate large-scale molecular line maps in a limited velocity range,
 especially in the $^{12}$CO and $^{13}$CO maps toward the Galactic center (e.g., see \cite{oka98,sof06}).  
 Furthermore, \citet{mol11} found a 100 pc twisted dust ring around the Galactic center on the dust temperature map created
 from 70 and 250 $\mu$m  dust emissions. Because emission from the foreground dark clouds must also contaminate this map,
 each foreground IR dust clump must carefully  be removed  from the map.
Studies of the infrared dark clouds with optically thin lines such  as NH$_3$ would be very useful 
for separating the foreground spiral arms from the Galactic center clouds.
 
\

The authors thank  Dr. Tomoharu Oka, Keio University, for useful comments. 



\begin{thebibliography}{}
\bibitem[Bae et al. (2011)]{bae11} Bae, J.-H.,  Kim, K.-T.,  Youn, S.-Y.,  Kim, W.-J.,  Byun, D.-Y.,  Kang, H.,  \& Oh, C. S.  2011, arXiv1108.3878 
\bibitem[Carey et al.(1998)]{car98} Carey, S. J., Clark, F. O., Egan, M. P., Price, S. D.,  Shipman, R. F.,  \& Kuchar, T. A. 1998, \apj, 508, 721 
\bibitem[Caswel et al.~(2010)]{cas10} Caswell, J. L.,  Fuller, G. A.,  Green, J. A.,  Avison, A.,  Breen, S. L. et al. 2010, \mnras, 404,1029 
\bibitem[Caswell~(1996)]{cas96} Caswell, J. L.  1996, \mnras,  283, 606 
\bibitem[Chambers et al.(2011)]{cha11} Chambers, Edward T., Yusef-Zadeh, F.,  \& Roberts, D. 2011, ApJ, 733, 42 
\bibitem[Chen et al. (2011)]{che11} Chen, X.,  Ellingsen, S. P.,  Shen, Z.-Q.,  Titmarsh, A.,  \& Gan, C.-G.  2011, \apjs, 196, 9 
\bibitem[Coudert \&  Roueff (2006)]{cou06} Coudert, L. H.\&  Roueff, E.  2006, \aap, 449, 855  [and  Erratum: \aap, 499, 347 (2009)]  
\bibitem[Cyganowski et al.  (2009)]{cyg09}	Cyganowski, C. J.,  Brogan, C. L., Hunter, T. R., \& Churchwell, E. 2009, \apj, 702, 1615
\bibitem[Danby et al.(1988)]{dan88} Danby, G.,  Flower, D. R.,  Valiron, P.,  Schilke, P.,  \& Walmsley, C. M. 1988, \mnras, 235, 229  
\bibitem[Deguchi et al.(2004)]{deg04} Deguchi, S., Imai, H., Fujii, T., Glass, I. S., Ita, Y., et al. 2004, \pasj, 56, 261 
\bibitem[Downes et al.(1982)]{dow82} Downes, D.,  Genzel, R., Hjalmarson, A., Nyman, L. A., \& Ronnang, B. 1982, \apjl, 252, L29
\bibitem[Dutra \&  Bica (2002)]{dut02} Dutra, C. M. \&  Bica, E. 2002, \aap, 383, 631  
\bibitem[Ellingsen (2005)]{ell05} Ellingsen, S. P. 2006, \mnras, 359, 1498 
\bibitem[Fontani et al.(2010)]{fon10}  Fontani, F.,  Cesaroni, R., \& Furuya, R. S. 2010, \aap, 517, 56 
\bibitem[G\"uver \&  \"Ozel~(2009)]{guv09}  G\"uver, T. \&  \"Ozel, F. 2009, \mnras, 400, 2050  
\bibitem[Haschick et al. (1990)]{has90} Haschick, A. D.,  Menten, K. M.,  \& Baan, W. A. 1990, \apj, 354, 556 
\bibitem[Jackson et al.(2008a)]{jac08a} Jackson, J. M., Finn, S. C., Rathborne, J. M.,  Chambers, E., T. \&  Simon, R. 2008, \apj, 680, 349  
\bibitem[Jackson et al.(2008b)]{jac08b} Jackson, J. M.,  Chambers, E., T.,  Rathborne, J. M.,   Simon, R. \& Zhang, Q. 2008, ASPC, 387, 44
\bibitem[Liszt et al.(2006)]{lis06} Liszt, H. S, Lucas, R., \& Pety, J. 2006, \aap, 448,  253 
\bibitem[Marshall et al. (2009)]{mar09} Marshall, D. J., Joncas, G., \&  Jones, A. P. 2009, \apj, 706, 727 
\bibitem[Martin-Pintado et al. (2000)]{mar00} Martin-Pintado, J., de Vicente, P., Rodríguez-Fernández, N. J., Fuente, A. \& Planesas, P. 2000, \aap, 356, L5
\bibitem[Menten~(1991)]{men91} Menten, K. M. 1991, in Atoms, ions and molecules: New results in spectral line astrophysics, ed. A. Haschick, \& P. T. P. Ho (San Francisco: ASP), ASPC, 16, 119
\bibitem[Molinari et al. (2011)]{mol11}  Molinari, S.,  Bally, J.,  Noriega-Crespo, A.,  Compi\'egne, M.,  Bernard, J. P. et al.  2011, \apj, 735, L33
\bibitem[Morris \& Serabyn~(1996)]{mor96}  Morris, M., \& Serabyn, E. 1996, ARA\&A, 34, 645
\bibitem[Nagayama et al.(2009)]{nag09}  Nagayama, T., Sato, S., Nishiyama, S., Murai, Y., Nagata, T., et al. 2009a, \pasj, 61, 283 
\bibitem[Oka et al.(1998)]{oka98} Oka, T., Hasegawa, T., Sato, F., Tsuboi, M., \& Miyazaki, A. 1998, \apjs, 118, 455
\bibitem[Oka et al.(2010)]{oka10}  Oka, T., Tanaka, K., Matsumura, S., Nagai, M., Kamegai, K., \&  Hasegawa, T. 2010, in "Proceedings of the Galactic Center Workshop" at Shanhai ( arXiv1002.1526)	
\bibitem[Pratap et al. (2008)]{pra08} Pratap, P., Shute, P. A., Keane, C.,  Battersby, C., \& Sterling, S. 2008, \aj, 135, 1718
\bibitem[Peretto \& Fuller (2009)]{per09}  Peretto,  N., \& Fuller,  G.A. 2009, \aap, 505, 405
\bibitem[Peretto \& Fuller (2010)]{per10}  Peretto,  N., \& Fuller,  G.A. 2010, \apj, 723,  555
\bibitem[Rathborne et al.(2006)]{rat06} Rathborne, J. M., Jackson, J. M., \& Simon, R. 2006, \apj, 641, 389
\bibitem[Rydbeck et al.(1977)]{ryd77}  Rydbeck, O. E. H., Sume, A., Hjalmarson, A., Ellder, J.  Ronnang, B. O., \& Kollberg, E. 1977, \apj, 215, L35
\bibitem[Schultheis et al.(1998)]{sch98} Schultheis, M., Ganesh, S., Simon, G., Omont, A., Alard, C., et al. 1999, \aap, 349, L69S 
\bibitem[Simon et al.(2006a)]{sim06a} Simon, R., Jackson, J.M., Rathborne, J.M., \& Chambers, E.T. 2006, \apj,  639, 227  
\bibitem[Simon et al.(2006b)]{sim06b} Simon, R., Rathborne, J. M., Shah, R. Y., Jackson, J. M., Chambers, E. T. 2006, \apj, 653, 1325 
\bibitem[Sofue~(2006)]{sof06} Sofue, Y.  2006, \pasj, 58, 335
\bibitem[Stolovy et al. (2006)]{sto06}  Stolovy, S., Ramirez, S., Arendt, R. G., Cotera, A., Yusef-Zadeh, F. et al. 2006, J. Phys. Conf. Ser, 54, 176
\bibitem[Taylor et al.(1993)]{tay93} Taylor, G. B., Morris, M. \& Schulman, E. 1993, \aj, 106, 1978 
\bibitem[Tsuboi et al.(1999)]{tsu99} Tsuboi, M.,  Handa, T.,  \& Ukita, N.  1999, \apjs, 120, 1 
\bibitem[Wang et al.(2006)]{wan06} Wang Y., Zhang Q., Rathborne J.M., Jackson J., \& Wu Y.  2006, \apjl, 651, L125  
\bibitem[Yusef-Zadeh et al.(2009)]{yus09} Yusef-Zadeh, F., Hewitt, J. W.,  Arendt, R. G.,  Whitney, B.,  Rieke, G., et al. 2009, \apj, 702, 178   
\end{thebibliography}
\end{document}